# THE ASSAY OF THE HYDRATION SHELL DYNAMICS ON THE TURNOVER OF THE ACTIVE SITE OF $CF_1$-ATPASE

*Alfred Bennun*[*]

Full Professor-Emeritus-Rutgers University
Consultant, CONICET, AR

## Abstract

Previous kinetic models had assumed that the reaction medium was reacting at random and without a turnover associated to thermodynamics exchanges, with a rigid active site on the enzyme. The experimental studies show that coupling factor 1 ($CF_1$) from spinach chloroplasts has latent ATPase activity, which become expressed after heat-treatment and incubation with calcium. The sigmoidal kinetics observed on the competitive effect of glycerol on water saturating a protein, suggests that the role of the hydration shell in the catalytic mechanism of the CF1-ATPase, modify the number of water molecules associated with the conformational turnover required for active site activity. It is assume that the water associated to the hydrophilic state of the enzyme produces a fit-in of the substrate to form (ES), follow by the catalytic action with product formation (EP). This one induces the dissociation of water and increases the hydrophobic attractions between R-groups. The latter, becomes the form of the enzyme interacting with water to form the dissociated free enzyme (E) and free product (P). Glycerol dependent suppression of the water dynamics on two interacting sites configuration shows a change in the H-bond-configuration. The thermodynamics modeling requires an energy expenditure of about 4kcal/mol per each H-bond in turnover. Glycerol determined a turnover of 14 molecules of water released from the active sites to reach the inactive form of the enzyme. The entropy generated by turnover of the fit-in and -out substrate and product from the protein could be dissipated-out of the enzyme-water system itself. Coupling with the surrounding water clusters allows to recreate H-bonds. This should involve a decrease in the number of H-bonds present in the clusters. These changes in the mass action capability of the water clusters could eventually become dissipated through a cooling effect.

[*] Corresponding author: Alfr9@hotmail.com

## Introduction

The coupling of ATP breakdown to ion-transport across the membranes when operate in reverse for ATP synthesis, is usually explained according to Mitchel's proton-motive force.

Experimentally has been probed that changing the coupling condition allows the same ATPase to participate in the catalysis of both phosphorylation and dephosphorylation paths. This indicates an eventual equilibrium excluding predominance of a one directional path. Effect explained according to microscopic reversibility. Conceptual justification is based in that a single open door as well as a single enzyme should allow transit in both directions at the same time.

Undoubtedly, an entropy decrease could not be predicted but is observed. The analog solution is to replace the common door for a rotatory one.
The ATPase structures itself, fits and provides for this revolving door action.

Experimental study of any supramolecular structures in bulk water is difficult because of their short lifetime: the hydrogen bonds are continually breaking and reforming at the timescales faster than 200 femtoseconds [1].

However, it has to be incorporated into the kinetic model, in order to comprehend the modulatory plasticity of the protein structure saturated by water.

Stoichiometry of the dynamics of the H-bonds sphere on a protein involves the H-bond balance on the surrounding water cluster required to sustain turnover.

The relationship $HO-H\ldots:OH^{+3}$ (18 kJ/mol or 4.3 kcal/mol; data obtained using molecular dynamics and should be compared to 7.9 kJ/mol for bulk water, obtained using the same molecular dynamics) [2].

The manner in which a solute modifies $H_2O$ within a mixing scheme depends on the nature of solute.

The H-bonds number fluctuates according to temperature [3]. Simulations at 25 °C indicated an average of 3.59 H-bonds per each water molecule.

## Methods

### Purification of CF1

Purification of CF1 (chloroplast coupling factor 1) was done by passage in DEAE-Sephadex column and equilibrated with 5 mM Tricine-NaOH, pH 8 and ATP to prevent denaturation of the coupling factor [4]. The eluated peak gave one major band with some tailing.

The desalted CF1 was transfer into 5 mM Tricine-NaOH, and 10 mM ATP solution, pH=8, to obtain a 20% glycerol-water v/v.

The mixture was subsequently lyophilized in a VirTis 10-010 freeze-dryer equipment to reduce to 1/5 of the original volume. The enzyme at concentrations of 1.5 to 4 mg of protein/ml in the water-glycerol solution stored at 2°C was stable over periods of about 6 months.

### Preparation of heat-Activated ATPase

Aliquots were diluted with 20 Tricine-NaOH, and 20 mM ATP pH=8, to decrease a final concentration to 0.5-2 mg/ml of CF1-protein in 10-25 (v/v) of glycerol-water.

The mixture was placed in a water bath at 65°C for 3 min to obtain the heat-activated ATPase. After this treatment disappears the coupling factor activity of CF1 and became exposed the function as an ATP hydrolytic site [5] [6] [7] [8] [9].

The heat-activated ATPase diluted with distilled water to final concentration of 30-150 μg of protein/ml, was assayed without delay for $Ca^{2+}$-stimulated ATPase activity.

Iodoacetamide treatment indicates that SH groups participate in the process of heat-activation, but not in catalytic activity [10].

### Assay

The final ATP concentration was computed within the final reaction mixture.

The pH was controlled in order not to deviate by more than 0.05 pH unit, from that reported for each experimental condition.

Released inorganic phosphate was determined by the method of Ernster et al. [11] or Nishizaki [12]. According to the procedure of Bernhardt and Wreath [13], it was measures the yellow color of the phosphomolybdic complex after its extraction with isobutanol-benzene.

Addition of acetone was omitted, but following removal of the water phase by suction, transparent clear isobutanol-benzene phases were obtained after centrifugation at 4,000 rpm for five minutes, and complete removal of residual water. The color was read at 320 mμ in a Zeiss PMQII spectro-Photometer. No ATP interference was observed. Calibration curves were linear up to 1.8 O.D. units.

The assay mixtures were incubated for 10 min at 37°C. Released inorganic phosphate was measured by the method of CHEN et al. [14]. The colorimetric reaction la highly inhibited by glycerol concentrations in excess of 6% v/v in the final volume of the colorimetric mixture. The colorimetric mixtures were adjusted to a final 3% glycerol concentration. Zero-time blanks were done showing that values were 2% less than that obtained in the absence of glycerol.

Other assay procedures by measuring the $^{32}P$ release from $AT^{32}P$. [14] were used to control the reproducibility of the experiments reported here.

The slope of the Hill curves (n) is accepted as the coefficient for interaction in the kinetic equation: $\log [v/(V_{max}-v)] - n \log [S] - \log K$ [15].

## Results

**Hydration shell dynamics on the active center of ATPase**

The Hill plot in Fig. 1b shows ATPase activity as a function of a glycerol titration competing with the hydration shell structure of the active site. The slope of the curve obtained has a value showing the displacement of 16 molecules of water from the active site for complete inhibition.

Therefore, the value kinetically attributable to the number of water binding sites in the enzyme is considerably larger than the two binding sites found for the substrate at the higher ATP concentrations.

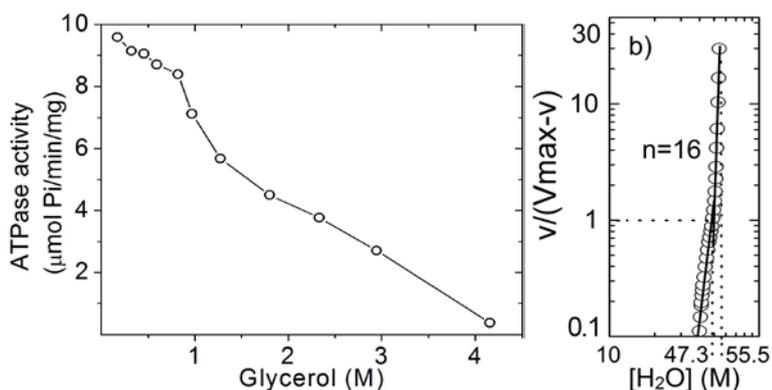

Fig. 1a: inhibitory effect of glycerol on heat-activated ATPase. 3.7 µg of heat-activated ATPase were incubated in 1.25 ml solution of 50 mM Tricine-NaOH, 8 mM ATP-CaCl$_2$ and varying concentrations of glycerol at the final pH 8. Fig. 1b: Hill's plot, as a function of the glycerol effect to decrease the molarity of water in the reaction mixture.

It may be that the difference of 14 water specific sites on the enzyme may reflect the state of solvation required to change the enzyme structure to an optimal catalytic conformation. This may be reflected in the finding that when the Hill plot is done as a function of glycerol concentration, the value of "n" equals 2. This value may indicate that the effect of glycerol solve for determining the number of subunits on the enzyme.

Sigmoidal inhibition by glycerol shows 50% inhibition of ATPase activity by 1.6 molar of glycerol, decreasing by 12% concentration of water in the reaction mixture indicates multiple and cooperative binding sites for water on the enzyme.

Changing the structure of the enzyme to a spontaneous hydrophobic conformation, involves less water for its state of solvation.

**The effect of the same concentration of glycerol at different pH**

The possibility that altering the state of solvation would result in changing the enzyme was investigating and reveals a change in pKa.

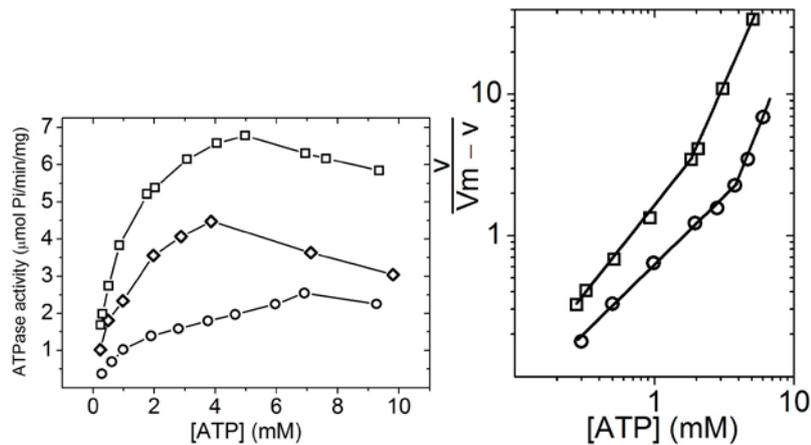

Fig. 2a: Effect of pH on the saturation kinetics of heat-activated ATPase. 4.3 µg of heat—activated ATPase were incubated in 1.25 ml solution of 50 mM Tricine-NaOH and varying equimolar concentrations of ATP and CaCl2 at the final indicated pH values. The assay mixtures were incubated for 7 min at 37oC. The reaction was terminated by addition of 0.1 ml of 20% trichloroacetic acid. Aliquots were analyzed for Pi [12]. Fig. 1b: Hill's plot.

Sigmoidal kinetics becomes more readily apparent at pH 7. Ideally in a Hill's plot the curves for the rate of reaction as the function log [v/(Vmax-v)] versus log [S] are linear if the reaction will follow a single order at all substrate concentrations.

Figure 2 illustrates studies of saturation kinetics at different pH values. The curve at pH 8.5 could be mistakenly thought of as following a hyperbolic function through all non-inhibitory substrate concentrations.

When the values shown in Fig. 1a are recalculated as illustrated in the Hill plot of figure 1b, the values obtained follow a curve that closely corresponds to the intersection of two lines with different slopes.

At substrate concentrations lower than 2 mM ATP (pH 8.5), or slightly higher (pH 7), the value of "n" is close to one. At all the different pH values studied "n" increases to two at the higher but non-inhibitory ATP concentrations.

Thus "n", the strength of the interaction between substrate binding sites appears to be modified at the higher substrate concentrations. Therefore, the apparent dissimilarity between the shapes of the curves at pH 7 and pH 8.5 as observed in Fig. 2a do not indicate that a pH change induces a transition of the reaction to another kinetic order.

At both pH values the transition occurs not as a pH effect but as a result of substrate cooperativity. As illustrated in Fig. 2b at pH 8.5, one half of the maximal velocity is reached at $7\times10^{-4}$ M, while at pH 7 the Km is $1.7\times10^{-3}$ M.

Therefore, a change of pH from 8.5 to 7 causes both Vmax and the affinity for the substrate (Km) to decrease without modification of the Hill curves. The effect of pH on the kinetics of the enzyme in water could be regarded as consistent with a model of mixed inhibition. This indicates that the enzyme may exist as equilibrium between more than one form, this equilibrium being displaced or regulated by protons [16].

**Glycerol induced near disappearance of the pH-dependent change**

The effect of the same concentration of glycerol at different pH values as a function of substrate concentration is illustrated in Figure 3a.

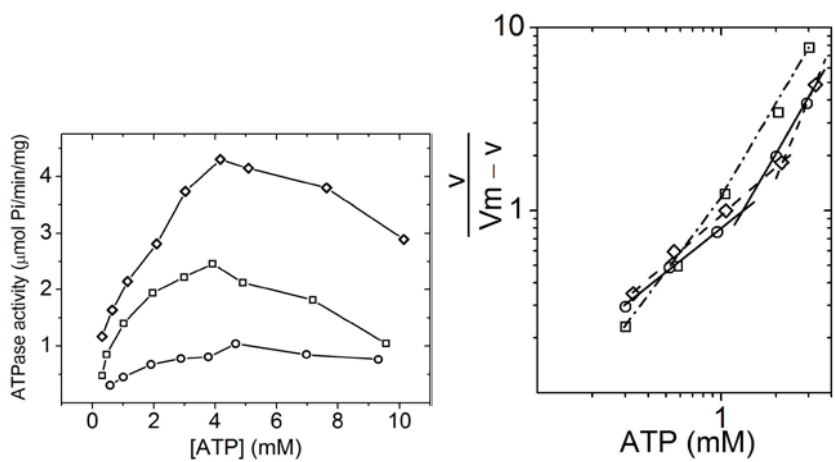

Fig. 3a: Effect of pH on the saturation kinetics of heat-activated ATPase in a glycerol-H2O mixture. Assay the incubating conditions and reaction mixture except for the addition of glycerol to a final 1.1 molar concentration (8% v/v), were identical as those described for the controls in the absence of glycerol reported in Fig. 2a. Inorganic phosphate was measured as described in Fig. 2a except that the aliquots were adjusted to a final 3% glycerol content before extraction with isobutanol-benzene 1:1. This aqueous dilution vas required to avoid interference of glycerol in the color development. Fig 3b: Hill's plot.

The enzymatic activity at all the studied pH values was decreased by about 30% from that of the control (Fig. 2b). The corresponding Hill plot (Fig. 3b) indicates Km values of $1\times10^{-3}$ M at pH 8.5 and $1.1\times10^{-3}$ M ATP at pH 7.

The decrease in Vmax with practically no simultaneous decrease in Km indicates an effect of pH on the kinetics of the enzyme in glycerol-water mixture which is closer to a noncompetitive inhibition model than the kinetics of the enzyme in water.

The specific conformation of a protein in solution is determined by the number of groups available to establish H-bonds with water molecules. The glycerol-induced near disappearance of the pH-dependent change in Km may indicate that changes in solvation could have been simultaneous and necessary for the pH-induced changes in Km observed with water. The ultraviolet spectrum of $CF_1$ in aqueous solution is strikingly different from that of its solution in a glycerol-$H_2O$ mixture.

Chloroplasts show undetectable endogenous ATPase activity [17]. However, conditions have been described under which ATPase activity in chloroplasts can be induced by light [18] [19].

The kinetics of these ATPase reactions indicates light-triggered and light-dependent mechanisms for the release of ATPase activity. This has been interpreted as indicating that activation of the ATPases has requirements for respectively, long-lived and short-lived light-formed compounds, or conformational changes [8].

Photophosphorylation is modified under the condition required for light-dependent and light-triggered ATPases [9]. The affinity for ADP and GDP as substrates (Km) of the residual and modified photophosphorylation is identical to their affinity as competitive inhibitors (Ki) of light-requiring ATPases [9]. The participation of an enzyme with a single active center in both ATPases and photophosphorylation is apparently contradictory because under optimal conditions photophosphorylation appears as a unidirectional catalytic process [17] [19].

However, it was found that a single coupling factor is required for the simultaneous reconstitution of chloroplast's light-requiring ATPases and for photophosphorylation [9]. Characterization of the chloroplast's coupling factor-1 (CF1) revealed weak ATPase activity in the dark [9] which could be increased several fold by trypsin [20], heat [20] [4] or dithioerithrol [21] treatments.

Therefore, light-activation of ATPases may be a phenomenon related to a potential in $CF_1$ for modification in its activity and function. In an attempt to characterize the properties of $CF_1$ which may allow modification of enzyme activity, studies were done on water ligand-induced changes which modify the heat-activated ATPase activity of $CF_1$.

However, it also requires validation of physicochemical parameters, which allow encoding brain states, and their retrieval, at the molecular level [22] [23] [24]. An applicable mechanism has to restrict thermic equilibrium [25] [26] [27], which at the protein structure level, may result in a background random noise, equivalent to 0.7kcal/mol at $37^{o}C$.

## Discussion

H-bonds between dissolved solute molecules are unfavorable relative to H-bonds between water and the donors and acceptors [28]. H-bonds between water molecules have an average lifetime of $10^{-11}$ seconds, or 10 picoseconds [29].

Atomic structure of an $H^+$-coupled ATP-synthase functions as a membrane rotor lead to the proposition that $H^+$ may be bound to these structures in the form of a hydronium ion.

*Ab initio* molecular dynamics simulations of the binding site demonstrate that the putative $H_3O^+$ deprotonates within 200 femtoseconds. The bound proton is thus transferred irreversibly to the carboxylate side chain found in the ion-binding sites of all ATP-synthase rotors. This result is consistent with classical simulations of the rotor in a phospholipid membrane, on the 100-nanosecond timescale. The observed coordination geometry is shown to correspond to a protonated carboxylate and a bound water molecule. Binding and transient storage of protons in the membrane rotors of ATP synthases occur through a common chemical mechanism, namely carboxylate protonation.

Microscopic reversibility is conceptually described as when a door is open it allows molecular transit in both senses, but the $CF_1$-ATPase in an membrane has the dynamics to turn around in similitude with a gyratory door, for only one way transit.

Hydration energy is how much energy is released when the ions are surrounded by water molecules. The heat released for 14 molecules of water in stoichiometric interaction with the enzyme $CF_1$-ATPase of molecular weight 360,000 g/mol appears to escape detection sensitivity.

## Conclusions

Transmembrane ATPases are described as anchored within biological membranes to move solutes across the membrane, against their concentration gradient.

The sigmoidal kinetics observed on the competitive effect of glycerol vs water saturating a protein, suggests that the role of hydration shell in the catalytic mechanism of the $CF_1$-ATPase is dependent of the modification of the number of water molecules acting on specific sites on the enzyme.

This allows for the fit-in of the substrate (ES) and transformation in unstable water-dependent state of association of the protein with the product (EP). Accordingly, some enzymes could be characterized by turnover between a hydrophilic form and hydrophobic one, providing through H-bond association vs dissociation states, the energy required to recover the original quaternary structure even after many catalytic cycles.

The ATPase rotatory movement within the membrane requires a single covalent form of the protein structure to have a site for binding substrate, but amenable to H-bonding changes at both sides of the membrane.

Mitchel's proton-motive force produces proton-translocation across membrane. A model molecular mechanism could be based in oxyHb, pKa=6, release of $O_2$ produces deoxyHb, pKa=8.2. The pKa differential involves that hydrophilic groups of low pKa are exposed allowing extensive proton-dissociation. The relax form (R) became favored because the dissociated acid groups could attract divalent metals to conform chelated structures.

The proton mass action reverses process displacing metal and passage to deoxy- form, increasing proton-association and pKa. The associated proton reduces the tendency to displace negative charges favoring instead hydrophobic attraction between R-groups in a tense form (T) as described for deoxyHb [30].

At the level of membrane processes the pKa differential between ATPase proteins in turnover, between hydrophilic and hydrophobic forms, could generate flows, supporting high vs a low proton pools. Similarly, at electrogenic membranes, changes in affinity for water could determine the transmembrane changes in Na$^+$ vs K$^+$ concentrations supporting action potentials.